# Bi Incorporation and segregation in the MBE-grown GaAs-(Ga,Al)As-Ga(As,Bi) core-shell nanowires

*Janusz Sadowski[1,2,*], Anna Kaleta[1], Serhii Kryvyi[1], Dorota Janaszko[1], Bogusława Kurowska[1],*

*Marta Bilska[1], Tomasz Wojciechowski[1,3], Jarosław Z. Domagala[1], Ana M. Sanchez[4], and*

*Sławomir Kret[1,*]*

[1] Institute of Physics Polish Academy of Sciences, Aleja Lotnikow 32/46, PL-02668 Warsaw, Poland

[2] Department of Physics and Electrical Engineering, Linnaeus University, SE-39182 Kalmar, Sweden

[3] International Research Centre MagTop, Institute of Physics, Polish Academy of Sciences, Aleja Lotnikow 32/46, PL-02668 Warsaw, Poland

[4] Department of Physics, University of Warwick, Coventry CV4 7AL, United Kingdom

**Abstract.** Incorporation of Bi into GaAs-(Ga,Al)As-Ga(As,Bi) core-shell nanowires grown by molecular beam epitaxy is studied with transmission electron microscopy. Nanowires are grown on GaAs(111)B substrates with Au-droplet assisted mode. Bi-doped shells are grown at low temperature (300 °C) with a close to stoichiometric Ga/As flux ratio. At low Bi fluxes, the Ga(As,Bi) shells are smooth, with Bi completely incorporated into the shells. Higher Bi fluxes (Bi/As flux ratio ~ 4%) led to partial segregation of Bi as droplets on the nanowires sidewalls, preferentially located at the nanowire segments with wurtzite structure. We demonstrate that such Bi droplets on the sidewalls act as catalysts for the growth of branches perpendicular to the GaAs trunks. Due to the tunability between zinc-blende and wurtzite polytypes by changing the nanowire growth conditions, this effect enables fabrication of branched nanowire architectures with branches generated from selected (wurtzite) nanowire segments.

Crystallization of Ga(As,Bi) ternary alloy is challenging since Bi acts as a surfactant during the epitaxial growth of III-V semiconductor layers[1]. However the application of dedicated thin film growth procedures allowed to obtain Ga(As,Bi) solid solutions with percentage range Bi content. Partial replacement of As anions by Bi heavy element leads to a substantial band-gap reduction and enhanced spin-orbit splitting in Ga(As,Bi), as compared with the binary GaAs host[2]. Ga(As,Bi) solid solution was first obtained by metalorganic vapor phase epitaxy (MOVPE)[3], and then by molecular beam epitaxy (MBE)[4]. The latter became the most conventional method of growing Ga(As,Bi) thin films, since higher Bi contents were obtained than those achieved with the use of the MOVPE technique[5]. GaBi binary compound does not occur in the crystalline form,[6] and therefore, the amount of Bi atoms residing in the As lattice sites in the GaAs crystal is limited. The highest Bi concentration in Ga(As,Bi) reported so far is about 20%, obtained in thin layers grown by MBE at optimized conditions. A significant energy gap reduction, down to 0.5 eV, has been reported for 17.8% Bi content[7]. Hence Ga(As,Bi) bandgap can be tuned between that of binary GaAs (1.43 eV at room temperature) and 0.5 eV, making this material suitable for infrared optoelectronic applications[8]. Moreover, Ga(As,Bi) and other heavily Bi doped III-V semiconductors (e.g. antimonides) are expected to exhibit topological insulator properties[9]. These topological properties were first based on theoretical modelling, but they have recently been experimentally demonstrated for InBi binary compound hosting topologically protected surface states[10,11]. There is no similar experimental evidence of topological protection in Ga(As,Bi) yet, but this ternary alloy has not been investigated in this context so far. Recently topological properties of Bi-alloyed III-V semiconductors (arsenides and antimonides) in the wurtzite (WZ) polytype have also been theoretically predicted[12]. Since WZ III-V compounds can easily be obtained in the quasi 1D (NW) geometry[13], the efforts to grow III-(V,Bi) ternary alloy NWs are interesting also in the context of topological materials. Although, Ga(As,Bi) planar layers grown by MBE have been investigated for over two decades, there is a very limited number of reports on Ga(As,Bi) in the NW geometry. The main reason is quite extreme MBE growth conditions of Ga(As,Bi) (in comparison with GaAs); i.e., low growth temperature (~ 200 – 350 °C) and close-to stoichiometric V/III elements flux ratio to avoid Bi surface segregation and to induce its substantial incorporation at As sites in the GaAs host lattice[14,15]. These growth conditions deviate from the GaAs NWs growth requirements, where higher substrate temperatures (> 500 °C) and high As excess are indispensable[16]. This implies that Ga(As,Bi) NWs with significant Bi content can only be obtained as low-temperature shells grown on Bi-free core NWs. This is similar to NWs implementing (Ga,Mn)As dilute ferromagnetic semiconductor with akin growth requirements, demanding even lower growth temperatures [~ 100 °C lower than Ga(As,Bi)] and the same  stoichiometric III/V flux ratio[17,18,19,20]. Recently the mixed phase WZ-ZB GaAs NWs exposed ex-situ to Bi vapor were studied by scanning tunneling microscopy[21] but this study has little reference to our investigations of  in-situ Bi incorporation.

In this paper we thoroughly investigate the radial distribution of Bi in GaAs NWs (not studied so far to our knowledge) and elucidate the emergence of side Bi droplets and branches on the wurtzite segments of GaAs NWs with Ga(As,Bi) shells.

## Samples and Experimental Methods

The core-shell nanowires were grown in a dedicated III-V MBE system. First the GaAs NW trunks were crystallized on GaAs(111)B substrates by the Au-assisted growth mode. 5 Å thick gold film was deposited on epi-ready GaAs(111)B wafers in another MBE system, and transferred (in air) to the III-V one. The growth was monitored by reflection high energy electron diffraction (RHEED) system. The substrate temperature was controlled by the MBE substrate manipulator thermocouple, calibrated using GaAs(100) surface reconstruction transition temperatures[22]. For that, a piece of GaAs(100) wafer was placed in the vicinity of Au-coated GaAs(111)B dedicated to the NWs growth. Planar Ga(As,Bi) layer grown on GaAs(100) during the Ga(As,Bi) NW shells deposition also serves as references to compare Bi incorporation into ZB and WZ GaAs phase of planar layers and NW shells, respectively (see the Supplementary Information). After preheating the substrates to 600 °C in the MBE growth chamber resulting in the thermal desorption of native oxide and formation of AuGa eutectic droplets at random surface sites of GaAs(111)B wafer, the substrate temperature was decreased to 540 °C, and GaAs NWs have been grown for 1-3 hours, depending on the sample. The Ga flux intensity (calibrated through a test growth on GaAs(100) substrate by RHEED intensity oscillations) corresponded to the planar growth rate of 0.2 μm/h, which resulted in the axial NWs growth rate of about 1 μm/h. As the source of arsenic, the valved cracker cell has been used with cracking zone temperature of 950 °C i.e., As dimers were prevailing in the As flux. The As/Ga flux ratio during this growth stage was about 10. The growth of GaAs NW trunks was completed by closing Ga and As shutters. The latter was closed one minute after the first one. After closing the Ga shutter the substrate temperature was decreased for deposition of the NW shells. In the case of one sample (sample 3) prior to the Ga(As,Bi) growth, about 30 nm (Ga,Al)As shells have been grown at the substrate temperature of 400 °C, with As and Ga flux ratios the same as used previously for the axial GaAs NWs growth. For the deposition of Ga(As,Bi) shells the substrate temperature was further decreased to about 300 °C. Ga(As,Bi) shells have been deposited in close-to-stoichiometric growth conditions i.e. with V/III flux intensity ratio close to 1.

We investigate three different types of GaAs-Ga(As,Bi) NW samples. The Ga(As,Bi) shells were grown in similar conditions for all the samples, i.e. at the same substrate temperature (300 °C) and As/Ga flux ratio (~ 1). The main difference between the growths was the Bi flux intensity during deposition of the low temperature (LT) Ga(As,Bi) shells. Bi flux was generated from a standard Knudsen effusion cell. To obtain different Bi concentrations the Bi cell temperature was set to 540 °C or 580 °C, corresponding to low (1%) or high (2-4%) Bi content in Ga(As,Bi) shells of sample 1, (2 and 3), respectively. In each case the Ga(As,Bi) shell was finished by the deposition of 4-7 nm thick LT GaAs.

Figure 1 shows the scanning electron microscopy (SEM) images of samples 1, 2 and 3 grown with low (sample 1) and high Bi flux (samples 2, and 3), at the Bi effusion cell temperature ($T_{Bi}$) of 540 °C and 580 °C, respectively. NW sidewalls in sample 1 have no side droplets. The conical features at the NW tips (with clearly visible Au droplet at the very top) are due to the residual axial growth during the Ga(As,Bi) shell deposition at 300 °C. In samples 2 and 3 grown with $T_{Bi}$ = 580 °C most of the NWs have side droplets, and short branches perpendicular to the main NW trunks.

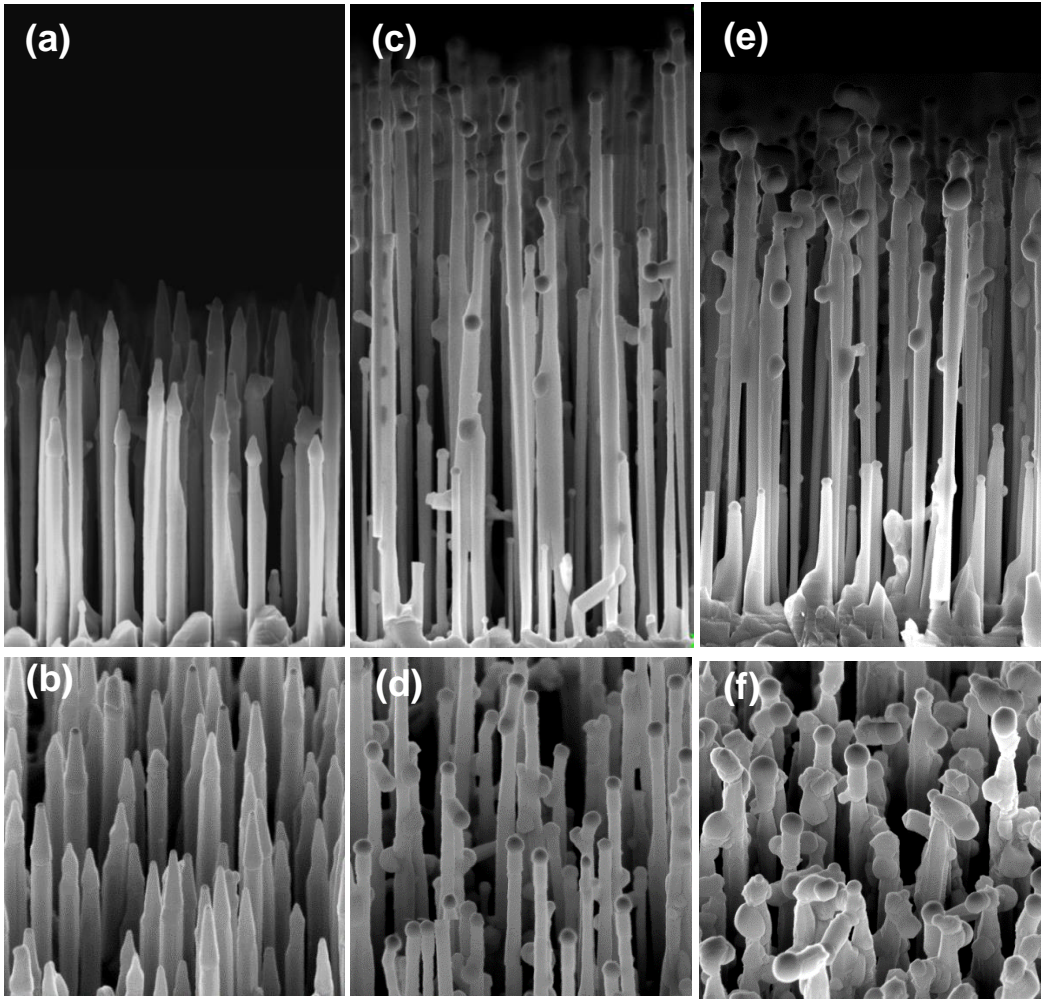

500 nm

**Figure 1.** SEM images of GaAs-Ga(As,Bi) core-shell nanowires grown on GaAs(111)B substrate. **(a,b)** sample 1 (1% Bi); **(c,d)** sample 2 (4% Bi); **(e,f)** sample 3 (4% Bi and additional (Ga,Al)As shell in-between GaAs NW trunk and Ga(As,Bi) shell. Upper panels – cross-sectional views, lower panels - 45° tilted views. The 500 nm scale bar plotted in the bottom part of the figure is common for all the panels.

A deeper insight into the NWs structure has been obtained by transmission electron microscopy (TEM) investigations. Morphology and structure of the NWs have been investigated using a FEI

Titan 80−300 transmission electron microscope (TEM) operating at 300 kV with a spherical aberration corrector of objective lenses in the HRTEM mode and a doubly corrected ARM200F microscope, operating at 200 kV. The elemental composition determination was carried out using EDAX 30 mm2 Si(Li) detector with a collection angle of 0.13 srad and a 100 mm$^2$ Oxford Instruments windowless EDX detector installed within the Jeol ARM200F microscope.

For TEM investigation the NWs have been transferred mechanically onto copper grids covered with holey carbon films enabling imaging the entire NWs. The cross-sections of the epoxy resin or platinum-carbon composite embedded NWs have been cut using a focused ion beam in Helios Nanolab 600 FIB as it is described in Ref.[23].

## Results and Discussion

Figure 2 shows TEM and scanning transmission electron microscopy (STEM) images of an individual NW from sample 1, grown with low Bi flux. Interestingly, pure ZB structure was revealed at the conical NW tip (Figure 2b and 2e). No distinct lines revealing stacking faults (SF) or twin boundaries (TB) are visible. We infer that these ZB tips emerge due to the residual axial growth during Ga(As,Bi) shell deposition. The pure ZB phase at the upper NW section is consistent with the reported dependence of GaAs NW phase on the NW tip–droplet contact angle[24]. The authors of Ref.[24] observed three different regions of droplet – NW tip contact angle ranges (here defined as β) inducing WZ or ZB structure during the NW growth. For low β values (80 − 100 deg.), the ZB phase is promoted, intermediate β values (100 − 125 deg.) promote WZ phase, and higher β values (125 − 140 deg.) again induce the occurrence of ZB phase in the NW. As can be observed in Figure 2b, the droplet contact angles for the NW tip are in the low β range, 90° on the left and 100° on the right side of the NW-droplet interface ($\beta_1$ and $\beta_2$, respectively), which according to Ref.[24] should promote ZB phase of GaAs NWs.

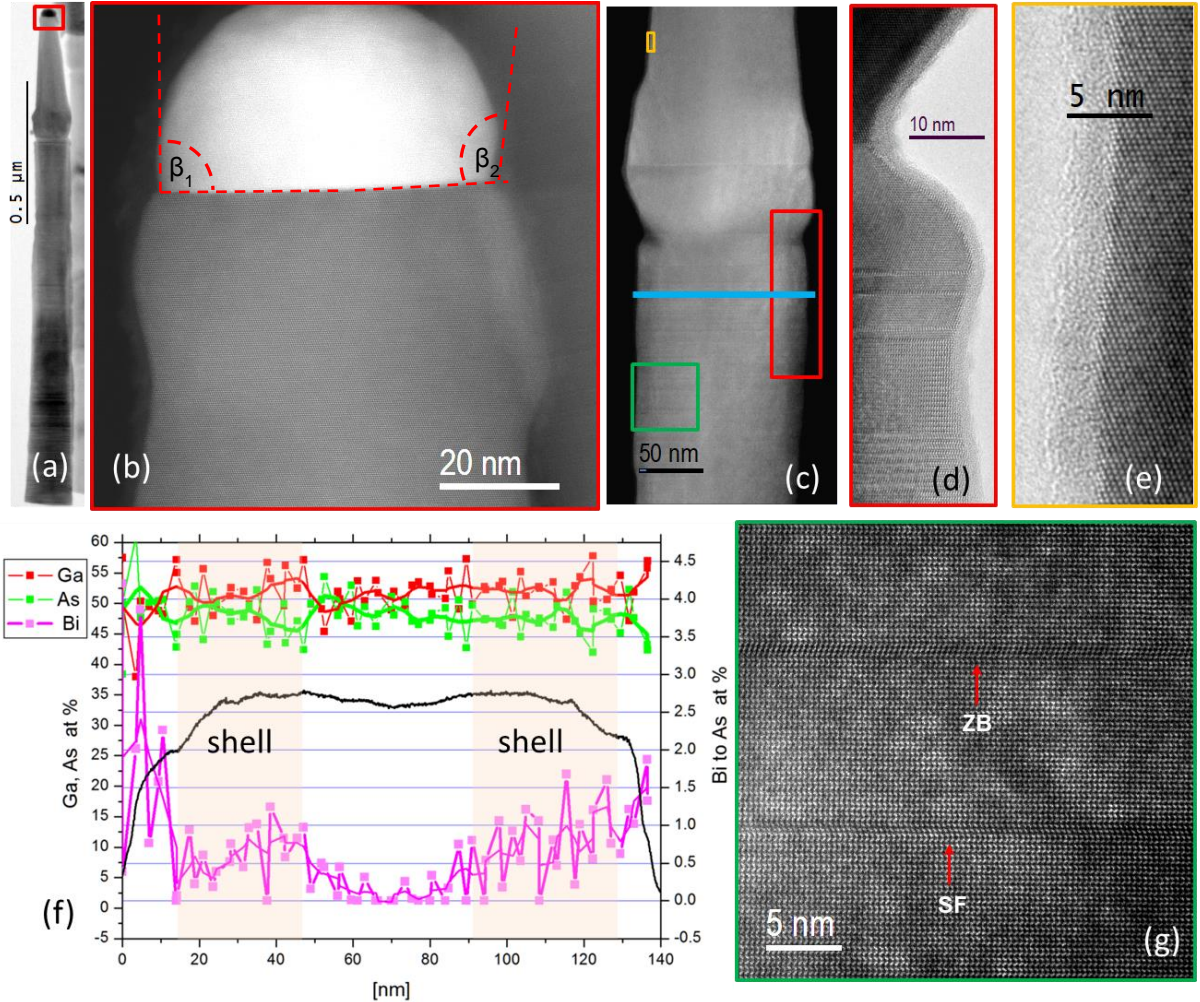

**Figure 2.** **(a)** TEM image of a GaAs-GaAs$_{0.99}$Bi$_{0.01}$ core-shell nanowire (sample 1), conical tip results from the residual axial growth during Ga(As,Bi) shell deposition; **(b)** ADF-STEM image of the upper zinc-blende stacking fault free part of NW below gold catalyzer droplet; **(c)** STEM image of the transition part of the NW between GaAs core and the section which grows during Ga(As,Bi) shell deposition; **(d)** HRTEM image showing progressive transition from WZ to pure ZB structure of the top NW part displayed in **(e)** – marked as small yellow rectangle in **(c)**; **(f)**

EDS elemental distribution profile for As, Bi and Ga along the blue line together with the STEM intensity; **(g)** typical structure of the bottom part of the NW bellow the transition zone evidencing mainly WZ structure, the single SF and the 3 ML thick ZB inclusion are also visible.

The Bi content values exceeding 1at.% at the surface of NWs (see the STEM intensity profile in Fig. 2f), are the quantification artefacts. Quantification is based on the Bi M-line of the intensity clearly above the noise level in the spectrum. Also the As concentration drops to 99 % with corresponding 1% Bi content detected. Quantification based on Bi L-line gives higher values of about 2 at.% but this line coincides with the As-K peak energy which can increase the apparent bismuth content due to not perfect deconvolution. The elemental distribution shown in Fig. 2f is expressed in at.%, for Ga and As, whereas the Bi concentration is normalized to the As atoms. From the Bi distribution and STEM-HAADF profile we can estimate the NW core diameter of ~ 40 nm and the Ga(As,Bi) shell thickness of ~ 30-40 nm. The maximum concentration of Bi is found at the NW side-wall surface (the amorphous "skin" of the NW – see Fig. S5 in the Supplementary Material), but the maximum concentration of Bi/As in the shell is at the level of ~1%. Interestingly in the planar ZB Ga(As,Bi) layer grown together with sample 1, the Bi content evaluated from the Ga(As,Bi) lattice parameter (see Fig. S1 in the supplementary material) amounts to 4.6%, which proves much effective incorporation of Bi into planar ZB GaAs(100) than to WZ GaAs(11-20) NW sidewall planes.

In order to increase the Bi content in Ga(As,Bi) NW shells, sample 2 was grown with much higher Bi flux intensity, than that used for sample 1. SEM images of sample 2 (Figure 1c,d) show longer NWs in comparison to sample 1 (Figure 1a,b). GaAs cores in sample 2 were grown for longer time (3 h 15 min vs 1h growth time relevant for sample 1), hence the NW lengths reach up to 4.5 μm. The Ga(As,Bi) shells were grown with $T_{Bi} = 580$ °C which corresponds to ~ 4 times higher Bi flux (sample 1 was grown with $T_{Bi} = 540$ °C). The detailed TEM analyses of the NW collected from sample 2 are shown in Fig. 3. The main difference between samples 2 and 1 is the occurrence of droplets at the NW sidewalls in the former – see Fig. 1c,d; Fig. 3a,d. Apparently, the Bi excess, which is not incorporated into Ga(As,Bi) shells, accumulates as droplets at the NW sidewalls. Additionally, a thin amorphous Bi layer can be observed at the NW sidewalls (Fig. S5 in the Supplementary Material). Moreover, as shown in Fig. 3d, a substantial Bi content is also revealed in the catalytic droplet at the NW top (solidified into nanocrystal at room temperature TEM investigations). The top droplet clearly consists of the two distinct parts. The part appearing as brighter seems to be embedded in the bigger, darker one (see Fig. 3d). The EDS analysis shows that the larger droplet consists of pure, solidified Bi. The elemental composition of the smaller part composition suggests that it contains a metastable $Au_{\sim0.40}Bi_{\sim0.60}$ phase[6]. The side droplet shown in Fig, 3a,b contains mainly Bi with a small admixture of As (up to 3 at.%). This droplet crystallized just below the transition region from mainly WZ to the pure ZB part. The details of this transition zone are shown in Fig. 3b,c. This WZ-ZB transition is quite abrupt, along about 5 nm long NW section. The upper ZB part is free form SF or TB defects. Figure 3d shows another NW also containing top and side droplet. Similarly to the first NW shown in Fig 3a,b, the top droplet also consists of two parts: pure Bi and Au-rich $Au_2Bi$ phase. Also in this case, the pure ZB structure part is about 100 nm long. The details of this part are shown in Fig. S4 of the supplementary material, revealing that it hosts an oblique SF which starts at the WZ-ZB transition zone and ends at $Au_2Bi/Bi$ interface. This suggests that $Au_2Bi$ phase nucleated at the SF and emerged during residual axial growth of this upper (ZB) NW part. Hence this part grew in a dual VLS and (vapor-solid-solid) VSS mechanism.

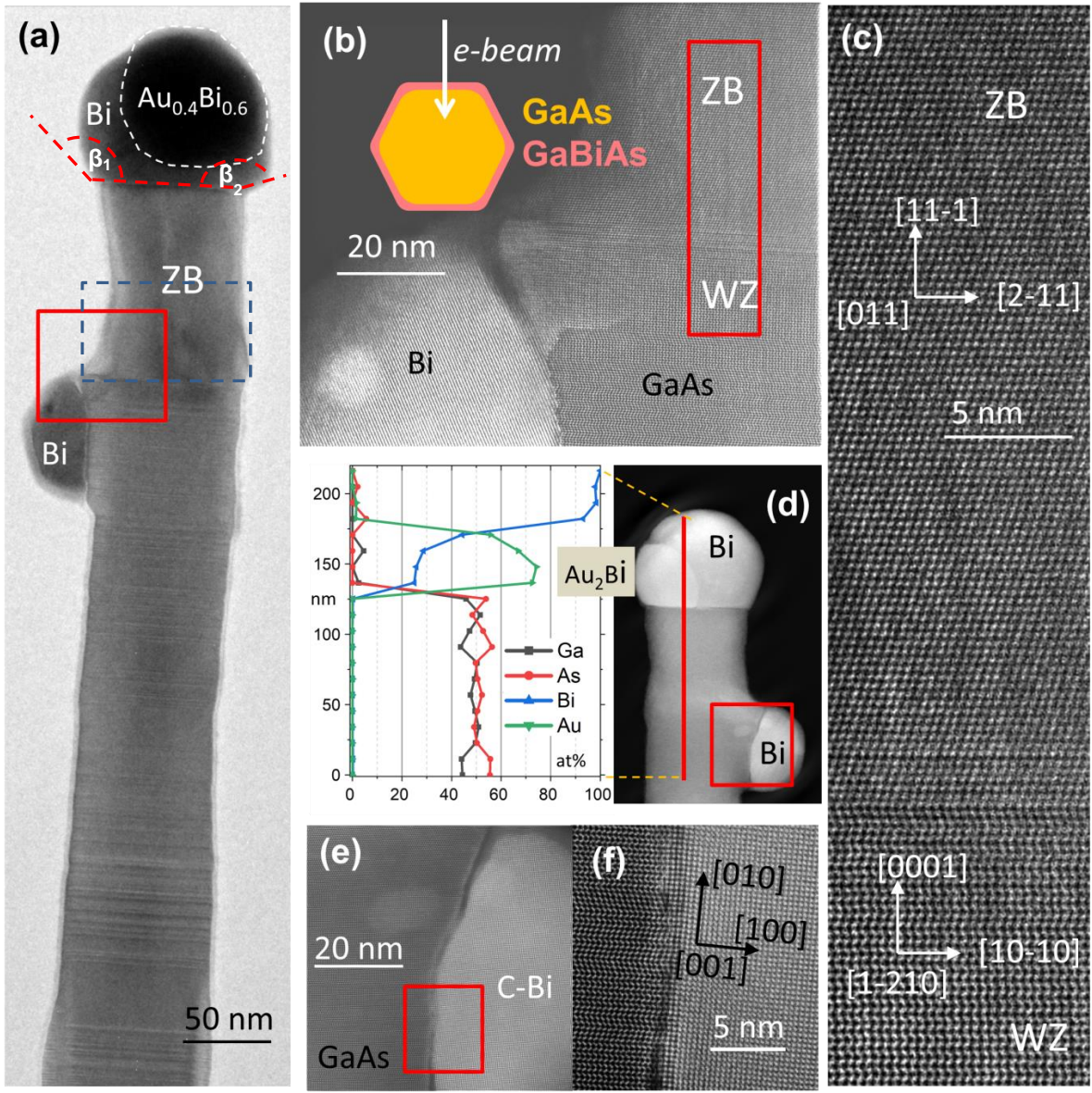

**Figure 3.** **(a)** HRTEM of a NW from sample 2; **(b)** STEM of the frame in **(a)**, a diagram shows orientation of the facets in relation to the e-beam; **(c)** STEM of the frame in **(b)**; **(d)** STEM image of another NW and EDS elemental concentration profile along the red line; **(e)** STEM zoomed part of the frame from **(d)**; **(f)** zoomed part of the frame marked in **(e)**.

The side Bi droplet crystallized in the uncommon cubic Bi phase[25] with the lattice parameter of 3.17 Å (see Fig. 3f). The (010) planes in the Bi droplet are in the epitaxial relation to WZ GaAs planes $d_{(0002)}$. The Bi lattice planes in vertical direction inside the droplet visible in STEM images (Fig. 3f) have the interplanar distance $d_{(010)}$=3.285 Å but in the radial direction the spacing $d_{(100)}$ equals to 3.33 Å. This means that the measured lattice parameter is about 3% bigger than that reported in Ref.[25] and we also detect a tetragonal distortion of 1.7%. The lattice planes of Bi in the (solidified) side droplets are rotated clockwise by 1deg. with respect to the core WZ GaAs lattice.

In contrast to sample 1 with pure top Au droplets with diameters of ~ 55 nm and 400 nm long, the conical ZB parts below the top droplet of sample 2, (see Fig. 3a) are much larger in diameter - about 100 nm, but the lengths of the top ZB segments are much shorter - about 110 nm. From now on, we will assign the name *"NW neck"* to the top ZB segment of the NW, which results from the residual axial growth during the low-temperature Ga(As,Bi) shell deposition (as discussed above). The shape differences of the NW necks in samples 1 and 2 can be explained as follows. In both cases, two growth modes of the NW neck occur simultaneously: (i) axial growth and (ii) radial growth. In sample 1, the catalytic droplet is small, without any Bi, since the Bi flux is low and all impinging Bi is incorporated into the NW shells (see Fig.1a,b and Fig.2a). It is well-known that when the growth is controlled by the diffusion flux of adatoms along the NW sidewalls towards the top, the axial NW growth rate is inversely proportional to the size of the catalytic droplet[26]. Smaller droplets in sample 1 induce faster residual axial growth during the Ga(As,Bi) shell deposition. The shell growth time in sample 1 was 1 hour, resulting in 400 nm long conical ZB NW neck. The Ga flux applied during the shell growth corresponded to the high temperature axial growth rate of the GaAs NW trunk equal to 1 μm/h. The axial growth rate of the NW neck during the 1 hour shell deposition is 2.5 times slower due to the low temperature of the shell growth (about 300 °C), and the competing mechanism of the radial growth. The conical shapes are typical for NWs grown at low temperatures, where the axial and radial growth modes occur simultaneously[16]. Here, the top droplet is ~100 nm in diameter; twice the size of the droplet in sample 1, but the ZB NW neck is only ~120 nm long (in contrast to 400 nm long neck of sample 1). The shell deposition time for sample 2 was 0.5 h (1h for sample 1), with the same Ga flux intensity, corresponding to a planar growth rate of 0.2 μm/h. The different neck shape in Sample 2 (in comparison to Sample 1) is most probably correlated to the larger top droplet size upon exposure to high Bi flux during the Ga(As,Bi) shell growth, with the top droplet gradually accumulating more Bi, which compensates the tapered NW shape (even reversed tapering can be inferred from Fig.3a induced by the low growth temperature). Standard NW tapering was observed in the first 50 nm of the NW neck (see section enclosed by the blue dashed rectangle in Figure 3a), whereas a slightly inverse tapering is visible in the section above. The common tapering effect associated with the radial growth, most pronounced at the lower part of the NW neck was compensated by the increasing diameter of the upper neck part, due to increasing size and Bi content of the top droplet. This effect can be observed in the two different NWs shown in Fig. 3a and 3d. The measured droplet contact angles $\beta_1$ and $\beta_2$ are equal to 135 and 200 deg. respectively in the NW tip - NW top droplet section shown in Fig. 3a. According to the in-situ TEM NW growth investigations reported in Ref. [24], the catalyzing droplet - NW tip contact angles higher than about 125 deg. promote the ZB GaAs NW phase.

To get better insight into the formation and morphology of Ga(As,Bi) shells, a ~15 nm thick (Ga,Al)As shells, with 30% Al, had been grown on the GaAs NW core before the Ga(As,Bi) shell growth, in the case of sample 3. The (Ga,Al)As shells were deposited at 400 °C. The lighter (Ga,Al)As layer contrast in ADF-STEM images (see Fig. 4b) allows to unequivocally identify the Ga(As,Bi) shells both in the plan-view and cross sections images of the NW. EDS compositional distribution across the WZ area of the NW shown in Fig. 4f reveals that the NW shell contains about 1% Bi.

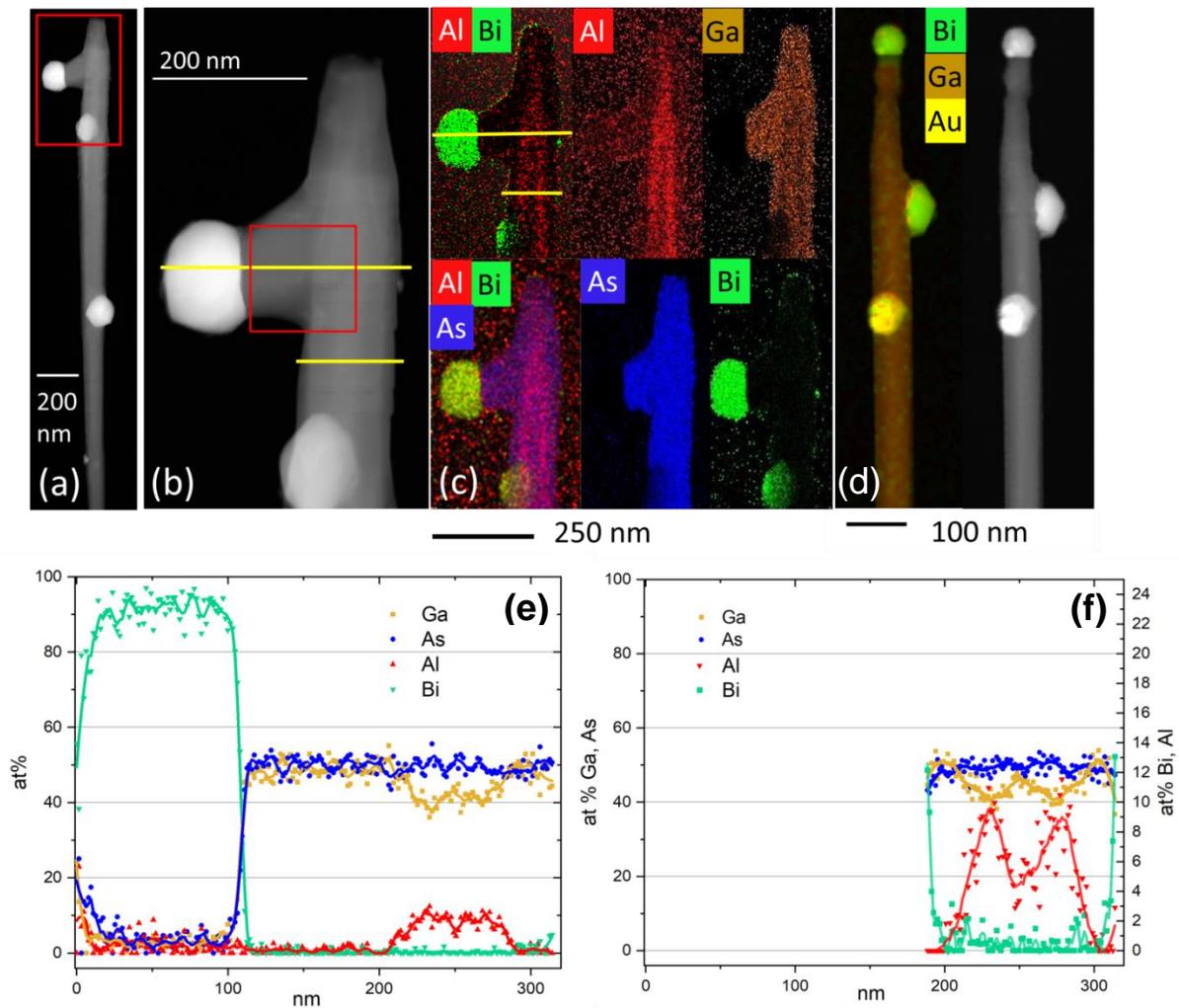

**Figure 4. (a,b)** TEM images of the upper GaAs-(Ga,Al)As-Ga(As,Bi) core-double shell NW part with Bi-catalyzed side branch (the very top part with the top catalyzing droplet has been broken during the NW collection), the side Bi droplet catalyzes the branch growth; **(c)** EDS composition maps visualizing the distribution of the elements: Ga, Al, As, Bi in the NW section shown in **(b)**; **(d)** – EDS composition map of another NW picked from Sample 3, showing Au droplet, which slid downwards and was replaced by the Bi droplet at the NW top; **(e,f)** – EDS line scans showing the distribution of the elements along yellow lines marked in **(b)** and **(c)**.

The Ga(As,Bi) shells were grown in the same conditions  (growth temperature, Bi and Ga fluxes, As/Ga flux ratio) as sample 2, but for a slightly longer time (45 min for sample 3, versus 30 min for sample 2). The 50% longer Ga(As,Bi) shell growth time under Bi excess conditions leads to a more pronounced Bi segregation forming more droplets at the NW sidewalls. Figure 4 summarizes the compositional analysis carried out for two representative NWs from sample 3. Two different upper GaAs-(Ga,Al)As-Ga(As,Bi) core-double shell NWs are displayed. The first one shown in Fig 4 a,b,c has a missing top. The HRSTEM images of the Bi droplet catalyzed side branch and branch-trunk regions are shown in Fig.S6 of the Supporting Information. Similarly to the case of sample 2 the residual axial growth during crystallization of Ga(As,Bi) shell produces distinct necks (visible also in Fig 1e, and 1f). This neck part can easily be broken during the mechanical transfer of NWs to the TEM grid. A complete NW is shown in Fig. 4 d and Fig. S7 in the Supporting Information. The EDS maps (Fig. 4c) and profiles (Fig. 4e,f) reveal the inner GaAs core and (Ga,Al)As shell. This data let us conclude that during the growth of (Ga,Al)As shell also axial (Ga,Al)As growth was continued over the length of  ~200  nm.

After the temperature drop necessary to grow Ga(As,Bi) shell, the gold droplet was replaced by the Bi one, and axial growth continued in both axial and slightly off-axis (random) directions, as can be seen in Fig.1e,f.

The accumulation of liquid Bi at the sidewall surface resulted in the formation of additional Bi-rich droplets, which started to catalyze the secondary branches reproducing the trunk structure - twinning ZB parts and SF in WZ parts of the core, as shown in Fig.S6 in the supporting information. Replacing the Au droplet by the Bi one at the NW top can be explained as follows. The temperature decrease to 300 °C for Ga(As,Bi) shell deposition caused crystallization of Au at the NW top, but after delivery of Bi in large amount, the external part of hitherto crystalized Au droplet became liquid. In the Au-Bi system, there are two eutectic points 371 °C and 241 °C, depending on the Bi concentration[6]. We infer that the 241 °C BiAu eutectic was formed and the surface of the gold nanocrystal at the NW top became liquid, which allowed the droplet to move and float down as shown in the EDS map in Fig. 4d. High magnification ADF-STEM image of the Bi-droplet catalyzed branch, corresponding to the region marked by the red square in Fig.4b, is shown in Fig.S6 in the Supporting Information.

Figure 5 shows STEM images and EDS composition maps of a 75 nm thick FIB cross-section of the ZB fragment of GaAs-(Ga,Al)As-Ga(As,Bi) NW (sample 3). High magnification ADF-STEM images are obtained using a HAADF detector with angular collection between $\alpha_{min} = 65$ mrad and $\alpha_{max} = 210$ mrad. The thickness of the analyzed NW cross-section is measured using position averaged convergent beam electron diffraction (PACBED) technique as described in the Supporting Information. Under these recording conditions, the HAADF images are sensitive to the average atomic number (Z) of the atom column (see Fig 5a). As can be seen in Fig. 5a the hexagonal shape of the GaAs core is reproduced in the successive shells. However, the last LT-GaAs outer shell does not develop sharp corners. The measured thicknesses of the (Ga,Al)As shells on all six NW sidewalls are equal to 18-20 nm.

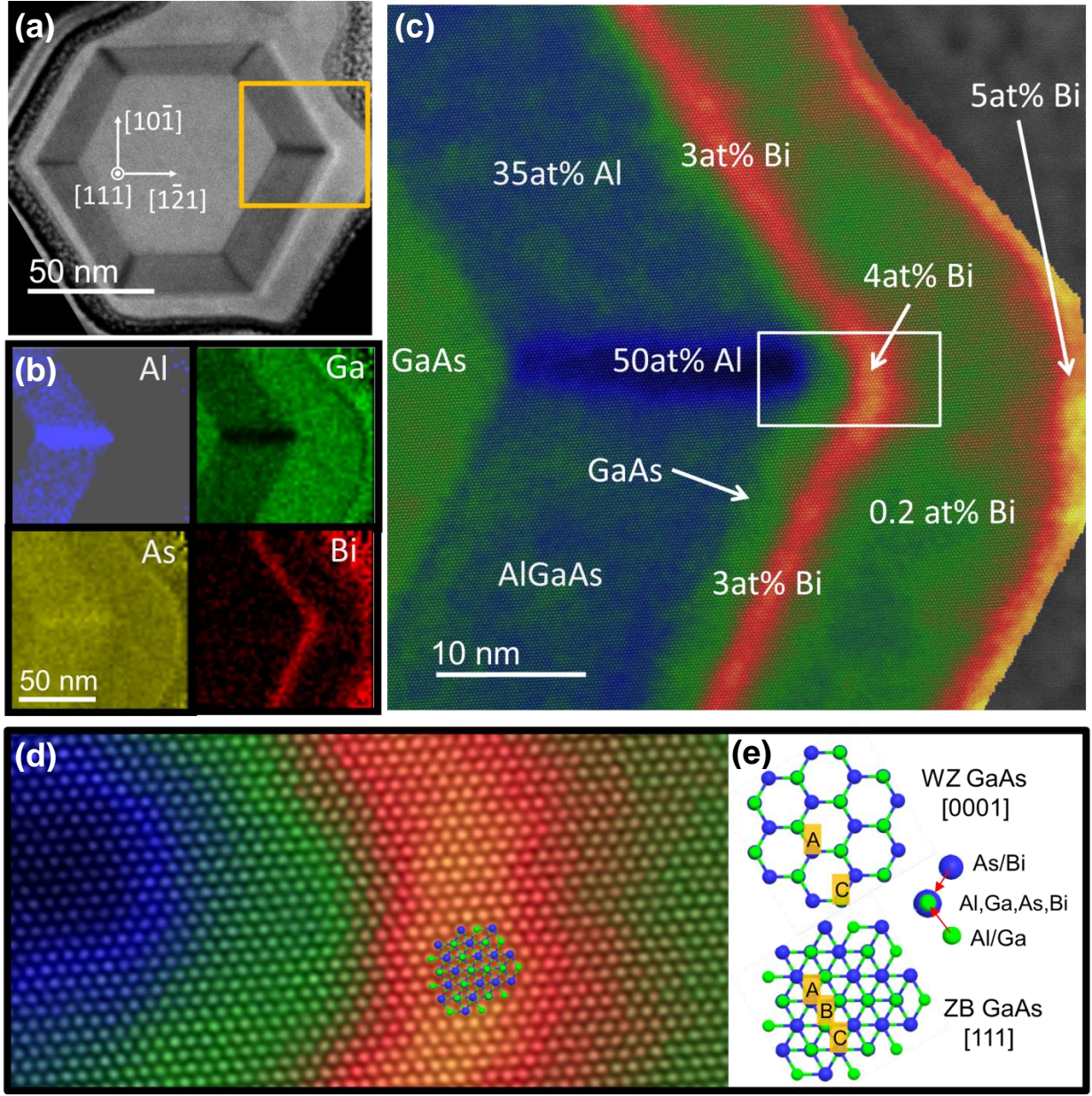

**Figure 5.** Analysis of a thin cross-section of ZB GaAs-(Ga,Al)As-Ga(As,Bi) NW segment.
**(a)** ADF-STEM image of the entire NW cross-section, the yellow frame indicates the area of EDS maps; **(b)** EDS distribution maps of individual elements (Al, Ga, As, Bi); **(c)** result of the semi-quantitative EDS analysis displaying Al, Ga and Bi distribution. Different colors and their saturation correspond to different composition of Ga, Al and Bi (green, blue and red, respectively) determined on the basis of the EDS signal and STEM intensity; **(d)** zoomed part of frame from **(c)** with superimposed  model of ZB sees towards  <111> ; **(e)** comparison of WZ and ZB  structure.

The dark lines running from the GaAs core corners visible in Fig. 5a have already been reported[27,28,29] for ZB NWs with (Ga,Al)As shells where high Al concentration, i.e. ~50% was detected in these regions and 30% Al elsewhere. However the Bi composition profile in the NWs with Ga(As,Bi) shells was not investigated so far, since no GaAs-Ga(As,Bi) NW cross-sections were studied before, to our knowledge. At the (Ga,Al)As-Ga(As,Bi) shell interface, the very thin adjacent Ga(As,Bi) shell region  with higher Bi concentration appears as brighter lines in Fig 5a (red stripes in Fig 5b and 5c). Moreover, additional, very thin bright lines are visible at the very edge of the NW (orange in Fig. 5c). It should be noted that the Ga(As,Bi) shell thicknesses are slightly different at each NW sidewall especially for the sidewalls in the closest neighborhood to other NWs, due to the shadowing effect during the MBE growth. Figure 5b shows the EDS elemental composition analysis of the cross-section specimen. The analysis was performed for the part enclosed by the yellow rectangle in Fig. 5a. The data presented in this Figure evidence direct correlation between the STEM image intensity and the concentration of elements visualized in the EDS maps as explained below.

The average intensity level of the STEM image in the GaAs core region amounts to 15800 counts per pixel what we consider as a reference value 1. STEM intensity reaches maximum of 19400 counts/pixel (1.23 in this convention) in the corners of hexagon which also corresponds to the maximum Bi concentration obtained from the EDS measurement. Moreover, in the areas of characteristic dark contrast (the features lying along the <112> crystallographic direction) the STEM intensity drops to 0.63.

Figure 5c shows a false-colored HR-STEM image obtained by combining information from EDS maps and intensities of STEM images. The STEM intensity was scaled based on the EDS results in the following way: in the case of Al and Ga the average ratio of Al/Ga from the shell region and from high Al content area were referred to the average STEM intensity in the same area. Then, the linear dependence of Al/Ga and STEM intensity was assumed. In the case of Bi, a similar procedure was applied but the average values for the middle of the Bi rich area were taken into account. For calculating the average chemical composition in the given area, the sum of spectra was used for elements quantification. Finally, the elemental composition maps were plotted with higher spatial resolution than original EDS maps of 50x50 pixels. The average Al concentration in the (Ga,Al)As shell is at the level of 30 at.% with respect to Ga. The maximum concentration of 50 at.% Al is reached at the corners of the (Ga,Al)As shell. The average concentration of Bi with respect to As in the NW cross-section reaches 3 at.% but it is higher (4 at.%) at the corners of the hexagon. The highest Bi concentration (5 at.%) is measured at the NW sidewall surface, however some Bi atoms are not embedded into GaAs lattice but remain at the surface as a very thin amorphous layer.  The average concentration of Bi between the Bi rich layer close to the inner (Ga,Al)As shells and the sidewall surface drops to  small values generally less than  1 at. %  (green areas in Fig, 5c).  However the Bi content in the areas with Bi concentration smaller than ~ 0.2 at% which is in our case under the detection limit (blue areas in green-colored Ga(As,Bi) shell) is only roughly estimated, however the intensity of Bi-M line in the average spectrum is still above the noise level. Additionally, between the (Ga,Al)As shell and Bi-rich shell, the pure GaAs region with a thickness ~ 1-2 nm is detected. Such a thin GaAs shell was intentionally deposited to avoid (Ga,Al)As surface contamination during the relatively long growth break required to stabilize a low temperature of Ga(As,Bi) shell deposition (300 °C) as well as to decrease and stabilize As$_2$ flux to close-to stoichiometric value to the Ga one. On other hand, the presence of such a layer clearly indicates that Bi bulk diffusion does not occur during the LT growth of the Ga(As,Bi) shell. The perimetric fluctuations of Bi concentration in the shell in the 2-5 nm scale are visible in Fig. 5c. However, averaging the information over 70-80 nm thick specimen blurs the localized Bi concentration fluctuations or clustering. Figure 5d shows atomic resolution STEM image of the area marked with the white rectangle in Fig. 5c, visualizing (Ga,Al)As-Ga(As,Bi) inner shell interface. The arrangement of atoms unequivocally points on the ZB structure of this NW slice. This is confirmed by the relevant ZB crystal structure model, shown in Fig. 5d and superimposed on the atomic resolution STEM image (Fig. 5c).

In Figure 5e, A, B, C refer to three positions of the cation/anion columns bonded in the ZB <111> or WZ <0001> direction (i.e. along the cation-anion bonding in the direction of tetrahedra heights). For a perfect crystal, without SFs, all the atom columns observed in the [111] or [0001] zone axis (ZB or WZ case, respectively) should correspond to the local maxima with the same intensities in STEM images, so they are indistinguishable. However, in the presence of SF in the WZ structure, contrast maxima in positions B can appear in the STEM image and local maxima can differ in intensities, which makes it possible to distinguish between ZB and WZ phase in a thin cross-section specimen.

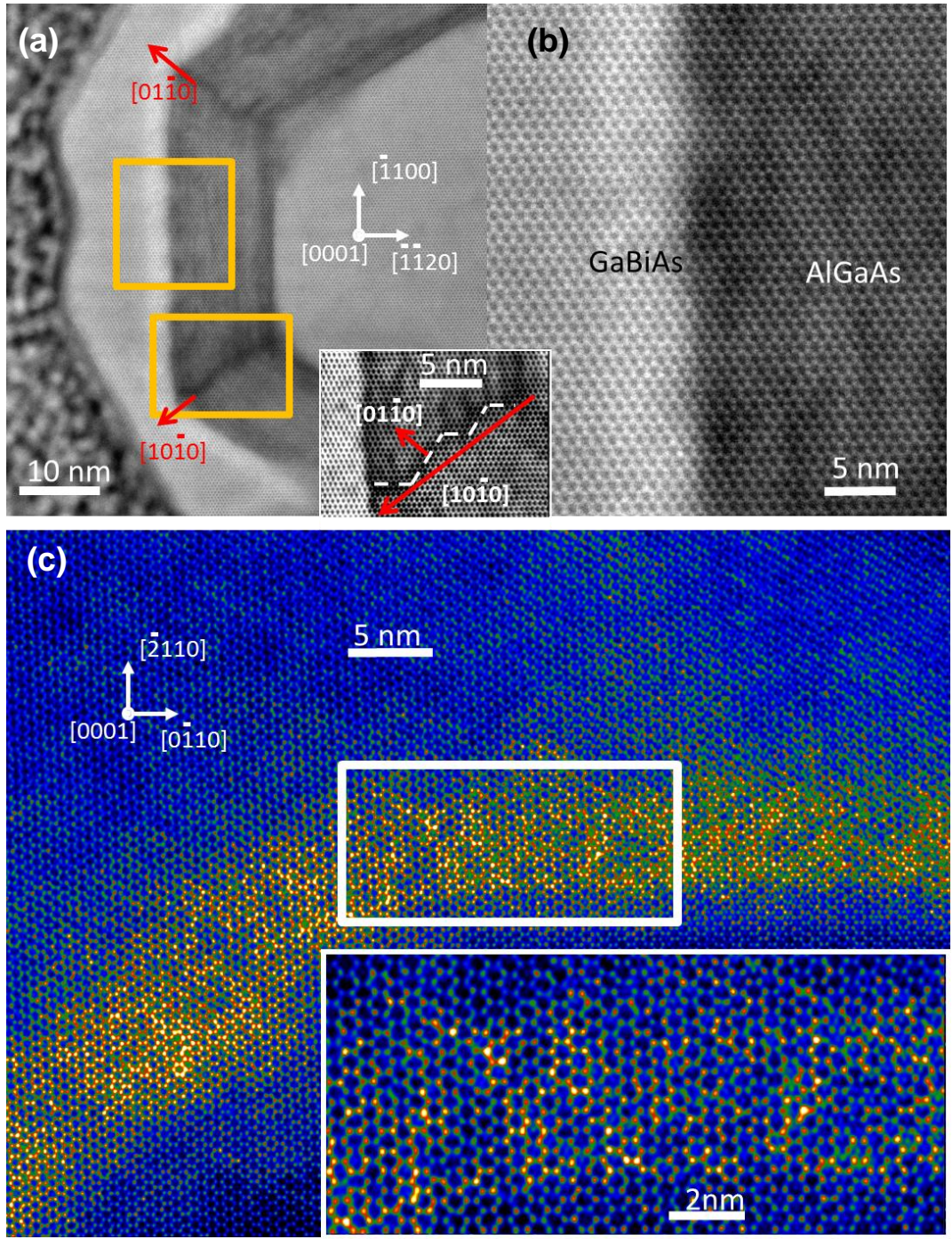

**Figure 6.** The analysis of a thin cross-section of the WZ GaAs-(Ga,Al)As-Ga(As,Bi) NW segment. **(a)** HR-STEM image acquired with the camera length of 73 mm; **(b)** zoomed part of the interface between (Ga,Al)As and Ga(As,Bi) shells; the inset to **(a)** and **(b)** - zoomed filtered image showing details of Al segregation at the NW corner; **(c)** HR-STEM (in false colors) of about 20 nm thick cross-section of Bi-rich shell; yellow color represents Bi-rich atomic columns averaged over 20-nm thick specimen, the inset shows zoomed area enclosed by the white rectangle.

Figure 6 shows a cross-sectional TEM image along the [0001] direction recorded in the WZ section of a NW from sample 3. In this case, the WZ phase can be easily identified due to fact that the only contrast maxima corresponding to …ACAC… stacking are visible, contrary to ZB part where maxima corresponding to ..ABC… stacking are visible (compare Fig 5). In this case, the hexagonal shape of the core is also preserved. More detailed discussion concerning distinctiveness of ZB and WZ phases in the TEM cross-section specimen is included in the supplementary material. The 5-6 nm thick Bi-rich shell can be unequivocally identified in Fig.6. This shell appears homogenous for all three NW facets visible in the image. We estimate that the specimen thickness is about 90 nm, so the fine Bi fluctuations cannot be detected. The segregation of Al is clearly visible inside the (Ga,Al)As shell. We also detect Al segregation at the edges of the hexagon, never reported before for WZ NWs. However, the segregation shown in the inset to Fig. 6a, 6b looks differently than that occurring in ZB (Ga,Al)As NW sections, reported earlier[27]. In our case, the high Al concentration regions have zigzag shapes[27]. Slightly similar zig-zag corner-line shapes were observed for P-rich regions of ZB Ga(As,P)/GaAs coaxial NWs[30]. The structural details of this Al-enhanced region are shown in the filtered image (inset to Fig. 6) where the zigzag shape, Al-rich path (dashed line) does not follow solely radial [10-10] direction but consists of two crystallographic directions [10-10] and [01-10]. This difference, with respect to the (Ga,Al)As NW shell of ZB structure is related to the lower symmetry of WZ structure, where the diffusion paths can be limited in comparison to the ZB one. The WZ NW cross-section shown in Fig. 6c is the thinnest specimen prepared with the FIB technique without its significant amorphization. We estimate the thickness of this cross-section to be less than 20 nm. Hence the local fluctuations of Bi atoms start to be visible with STEM and are much higher than those in the Bi-poor LT-GaAs shell. The Bi-rich areas are generally random but some texture in the radial direction can be distinguished; such areas with the diameters of 1-2 nm can be seen. In the cross-sections of both WZ and ZB parts of a NW collected from sample 3, we notice inhomogeneous concentration of Bi in the radial direction. Similar Bi concentration profile was also observed in the MBE-grown planar Ga(As,Bi) layers, with Bi concentration exponentially dropping along the growth direction at first 25 nm Ga(As,Bi) film thickness and then stabilizing at a constant level[31]. In the core-shell NW heterostructures investigated here, the thickness of Ga(As,Bi) shells is in the range of 14-26 nm, hence the same effect can occur. Moreover, as revealed by the Monte Carlo simulations of Bi incorporation and droplets formation during the MBE growth of Ga(As,Bi), once the surface Bi droplets are formed the incorporation of Bi into the growing film decreases considerably[15]. In our case, both effects can occur. In the cross-section of a ZB NW part, we also observe the enhanced Bi content at the outer NW region close to the sidewall surface and at the outer NW corner.

## Concusions

In conclusion, we have investigated Bi incorporation into predominantly wurtzite GaAs-(Ga,Al)As-Ga(As,Bi) core-shell nanowires grown by molecular beam epitaxy. At low Bi content of 1% the NWs are smooth, with no additional features at their sidewalls, whereas at the attempted Bi content of 4%, the droplets are formed at the NW sidewalls. At the NW tops Bi merges with the Au catalyst and induces (in the NWs cooled down to room temperature) the phase separation into $Au_2Bi$ and pure Au, however with undisturbed spherical top nanoparticles shapes. In the cross section of the core-multishell GaAs-(Ga,Al)As-Ga(As,Bi) NWs we have detected radially inhomogeneous distribution of Bi. The enhanced Bi content occurs at the inner interfaces of Ga(As,Bi) shells and on their outer surfaces (NW sidewalls). Additionally, in the WZ (Ga,Al)As shells we have observed, so far unnoticed zigzag-shaped enhanced Al content regions along the lines extending radially out of the hexagonal WZ NW corners. In the NWs with mixed ZB-WZ axial structure the preferential location of the side Bi droplets at the WZ NW sections was evidenced. Such Bi droplets induce the growth of WZ GaAs branches perpendicular to the NW trunks. The catalyzing action of Bi droplets, with preferential location at wurtzite nanowire segments, enables controlled formation of branches in the GaAs NWs with mixed WZ-ZB structure, since the structural change between both GaAs polytypes can be induced by alterations of the NW growth conditions.

**Supporting Information:**

XRD spectrum, SEM and TEM images of a planar Ga(As,Bi) layer grown together with GaAs-Ga(As,Bi) core shell nanowires (Sample 1); additional TEM images of NWs and NW cross-sections including EDS composition maps (Samples 2 and 3).

## Acknowledgements

This work is supported in part by the National Science Centre, Poland through grants 2016/21/B/ST5/03411, 2017/25/N/ST5/02942, 2018/31/B/ST3/03610; by the EAgLE international project (FP7-REGPOT-2013-1, No. 316014), and by the Foundation for Polish Science through the IRA Programme co-financed by EU within SG OP (Grant No. MAB/2017/1). The authors thank Dr. A Pietruczik from IP PAS, for deposition of Au on GaAs(111)B wafers.

## Author Contributions

J.S. grew the samples and designed the project. J.S., A.K., S.Kret. A.M.S contributed to the writing of the manuscript. TW performed SEM characterization, S.Kret., A.K., A.M.S S. Krivyj did TEM characterization and interpreted TEM results; B.K. M.B. D.J, contributed to preparation of TEM specimens. J. Z. D. did XRD measurements implemented in the Supporting Information file. All authors approved the final version of the manuscript.

# Supplementary material

# Bi incorporation and segregation in the MBE-grown GaAs-(Ga,Al)As-Ga(As,Bi) core-shell nanowires

*Janusz Sadowski[1,2], Anna Kaleta[1], Serhii Kryvyi[1], Dorota Janaszko[1], Bogusława Kurowska[1], Marta Bilska[1], Tomasz Wojciechowski[1,3], Jarosław Z. Domagala[1], Ana M. Sanchez[4], and Sławomir Kret[1]*

[1] Institute of Physics Polish Academy of Sciences, Aleja Lotnikow 32/46, PL-02668 Warsaw, Poland

[2] Department of Physics and Electrical Engineering, Linnaeus University, SE-39182 Kalmar, Sweden

[3] International Research Centre MagTop, Institute of Physics, Polish Academy of Sciences, Aleja Lotnikow 32/46, PL-02668 Warsaw, Poland

[4] Department of Physics, University of Warwick, Coventry CV4 7AL, United Kingdom

# 1. Planar Ga(As,Bi) layers grown together with GaAs-Ga(As,Bi) core-shell nanowires.

## 1.1. Characterization by X-ray diffraction and SEM

The planar layer grown on the Au-free GaAs(001) substrate together with sample 1 (GaAs – Ga(As,Bi) core-shell NWs) was measured by high resolution X-ray diffraction (XRD) X'Pert MRD with Cu tube (CuK$_{\alpha 1}$ radiation ($\lambda = 1.5406$ Å)), equipped with: X-ray mirror, monochromator (asymmetrically cut Ge 4x(220)) and with two proportional detectors, one of the detectors is preceded by the analyzer 3xGe(220).

Figure S1 shows results of 2θ/ω measurement around 004 Bragg reflection of GaAs(001) substrate.

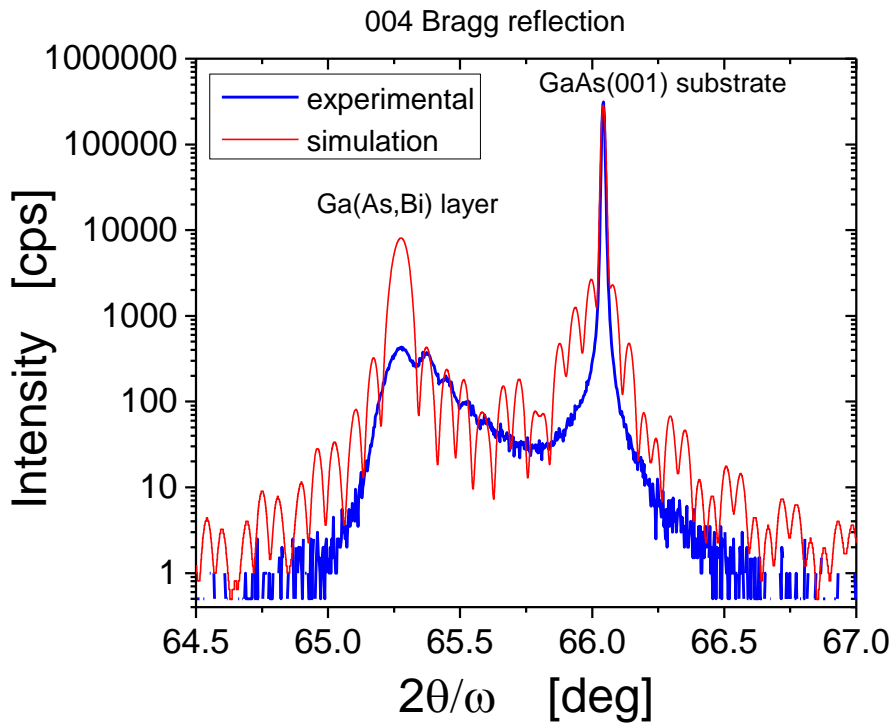

**Figure S1**. X-ray diffraction results - 2θ/ω measurement, 004 Bragg reflection of a planar Ga(As,Bi) layer grown together with the nanowires of sample 1, with Bi effusion cell temperature T$_{Bi}$ = 540 °C.

The perpendicular lattice parameter obtained from the angular position of the peak corresponding to Ga(As,Bi) layer (grown simultaneously with Ga(As,Bi) shells of GaAs NWs on Au-coated GaAs(111)B) equals to: $a_\perp$ = 5.7126 Å. The Ga(As,Bi) layer is fully strained to the GaAs substrate, hence we can calculate the relaxed lattice parameter (assuming the Ga(As,Bi) elastic constant values $C_{11}$ and $C_{12}$ to be the same as for GaAs) which amounts to: $a_{relaxed}$ = 5.68473 Å. Taking the hypothetical lattice parameter of binary GaBi equal to 6.33 Å and assuming Vegards law we obtain then 4.6% Bi content in the planar Ga(Bi,As) layer grown together with sample 1. The distance between pendelösung fringes, confirmed by the XRD simulations yields the thickness of Ga(As,Bi) planar layer equal to 137 nm, which agrees quite well with the value assumed from the GaAs growth rate calibrations based on RHEED oscillations.

Apparently, in the planar zinc-blende (ZB) counterpart of sample 1 the incorporation of Bi is much higher than that in the WZ Ga(As,Bi) NW shell. EDS signal for Bi is low and from the average spectra we obtain values at the level of 1at %, however with large error (about the same order of magnitude, i.e. ±1%). Nevertheless, the Bi-M line in the X-ray spectrum is clearly above the noise level. Analysis of different spectra gives similar results, around 1at%.

The surface of the reference planar layer grown together with sample 1 is a bit "milky" (as observed by naked eye), suggesting the presence of some 3D objects (Bi droplets) but it still shows distinct 2D RHEED patterns, hence the concentration of the droplets is not huge. However, for samples 2 and 3, grown with about 4 times higher Bi flux, corresponding to the temperature of Bi source $T_{Bi}$ = 580 °C (40 °C higher than that used during the growth of sample 1), RHEED images for reference planar layers disappeared completely upon the growth of Ga(As,Bi). This indicated the presence of amorphous surface features yielding diffused uniform RHEED background. This can be attributed to liquid Bi droplets (the growth temperature is equal 300 °C which is slightly higher than the Bi melting temperature 270 °C)). These droplets (solidified at room temperature of SEM measurements) are shown in Fig. S2, both in plan view and in cross-sectional sample orientation.

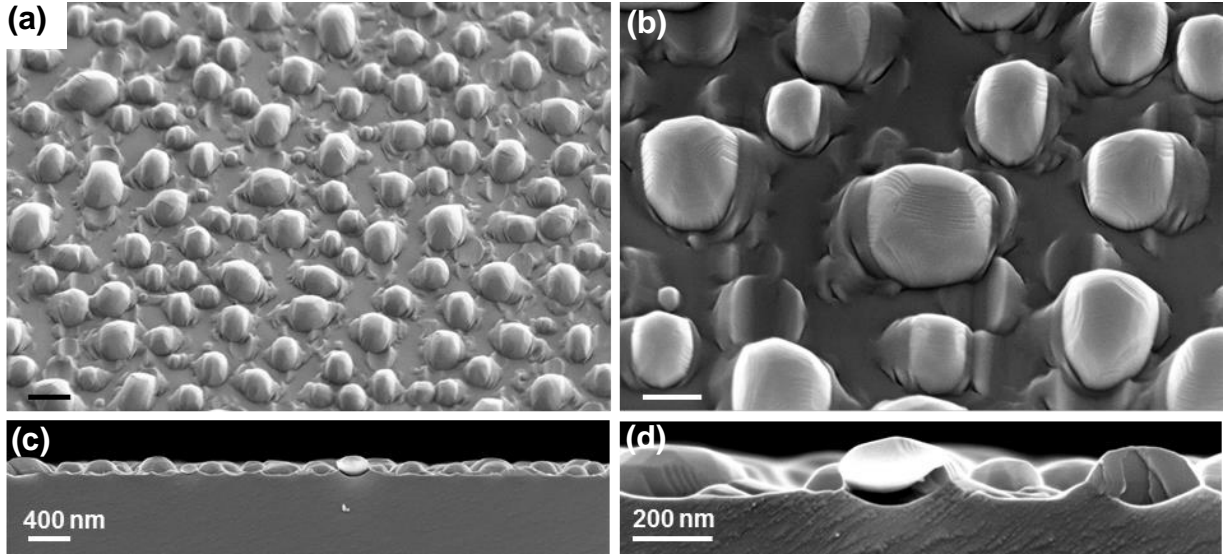

**Figure S2.** SEM images of the surface of Ga(As,Bi)(100) planar layer grown together with Ga(As,Bi)(111) nanowires (sample 2). Panels (a) and (b) - plan view; panels (c) and (d) - cross sectional view of the cleaved sample edge. The scale bars correspond to 400 nm in the panels (a), (c) and to 200 nm in the panels (c) and (d).

## 1.2. Characterization by cross-sectional TEM

More insight into the planar Ga(As,Bi) layer and individual surface Bi droplets is provided by the cross-sectional TEM image of a planar reference to sample 2.

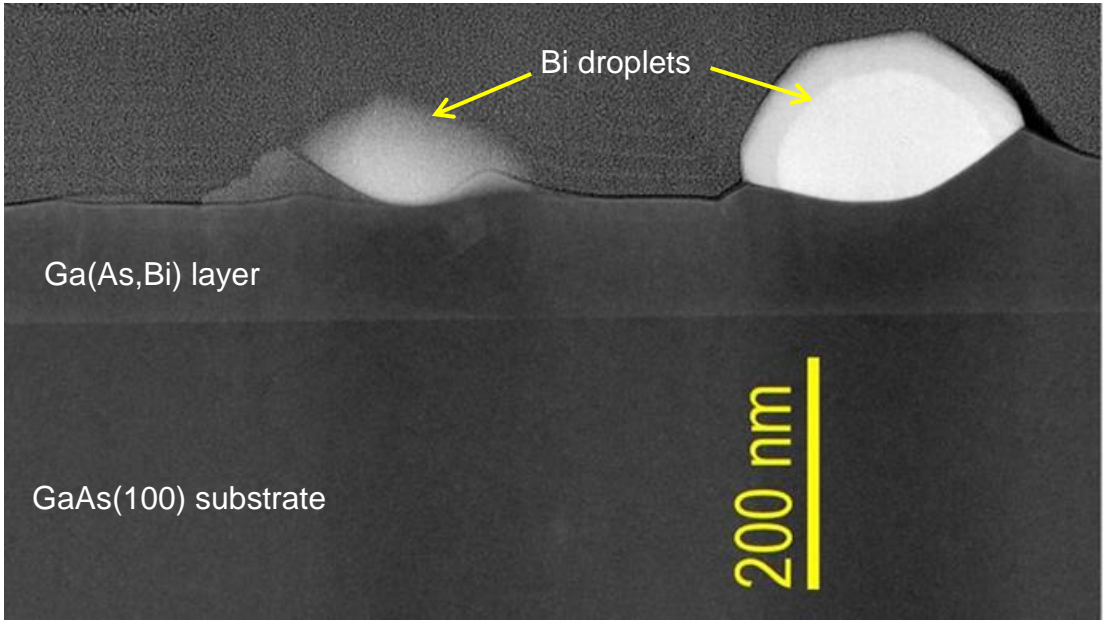

**Figure S3.** TEM cross-sectional image of Ga(As,Bi)(100) planar layer grown together with Ga(As,Bi)(111) nanowires (sample 2).

Interestingly the surface morphology of planar Ga(As,Bi) in-between segregated Bi droplets is consistent with the predictions based on the Monte-Carlo simulations of Bi droplet segregation during Ga(As,Bi) MBE growth.[1]

## 2. Supplementary information on nanowires

Figure S4 shows the top part (a) - ZB, (b) - WZ of the NW taken form sample 2. The oblique stacking fault (SF) defect visible at panel (a) is about 30 deg. inclined vs NW axis, and its end separates two distinct regions of the top droplet - $Au_2Bi$ and pure Bi. In sample 2 we could identify a lot of NWs with similar top parts.

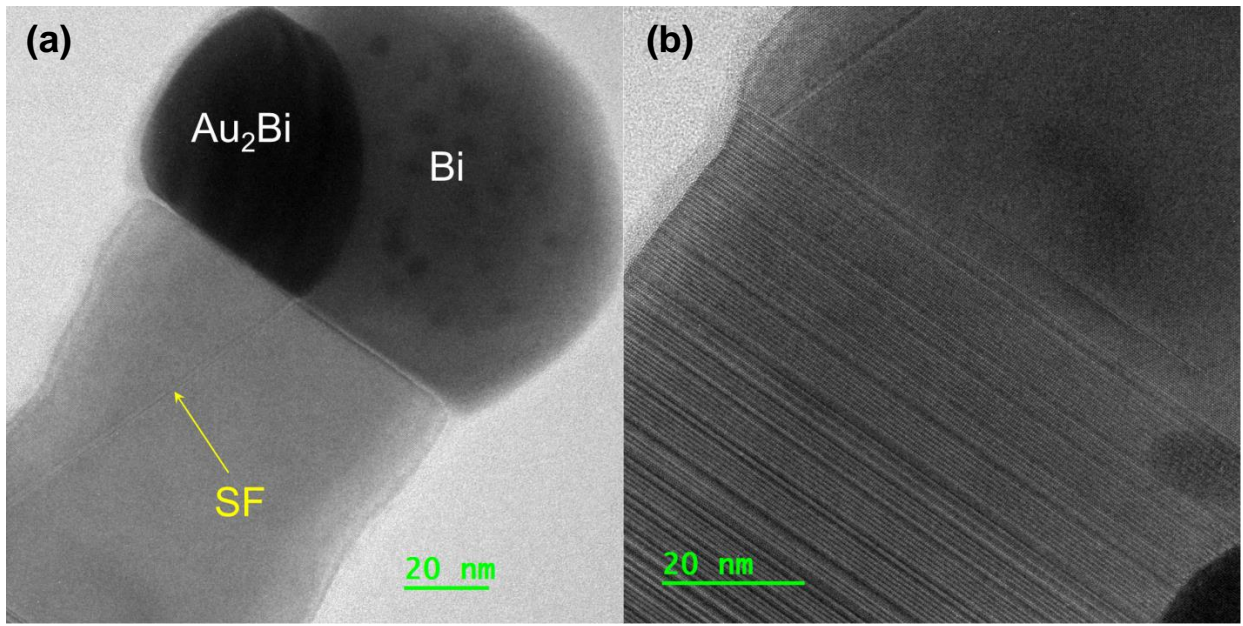

**Figure S4.** (a) - Top (ZB) part of a NW taken from Sample 2 showing inclined stacking fault whose end separates two distinct regions of the top droplet; Au$_2$Bi and pure Bi; (b) - The WZ part below the ZB NW neck contains a large number of typical SF defects, perpendicular to the NW axis.

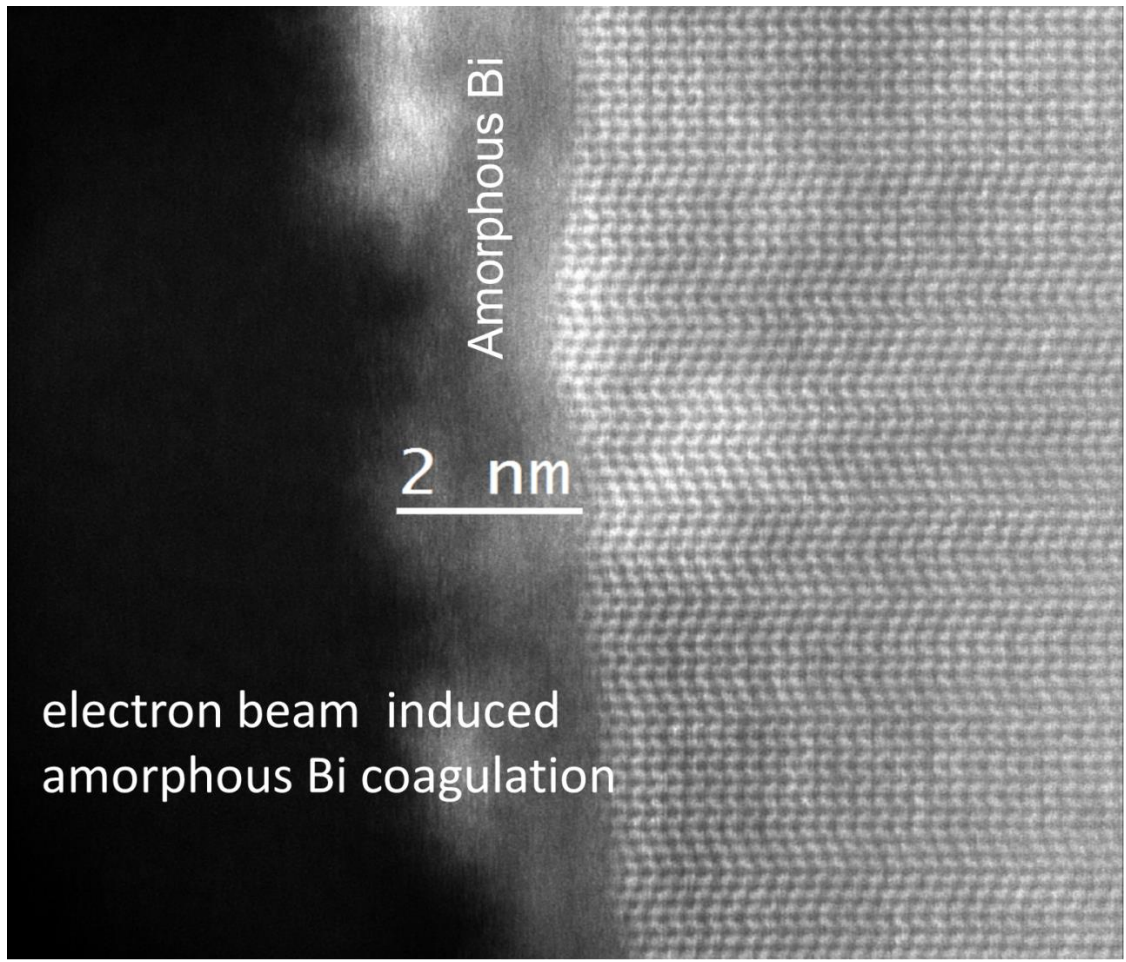

**Fig. S5.** Amorphous Bi "skin" on the sidewall of Sample 2. The Bi amorphization is induced by the e-beam during HRTEM examination.

Figure S.6 shows HRSTEM images revealing the details of the WZ branch generated at WZ NW trunk by the side Bi droplet in the NW taken from sample 3.

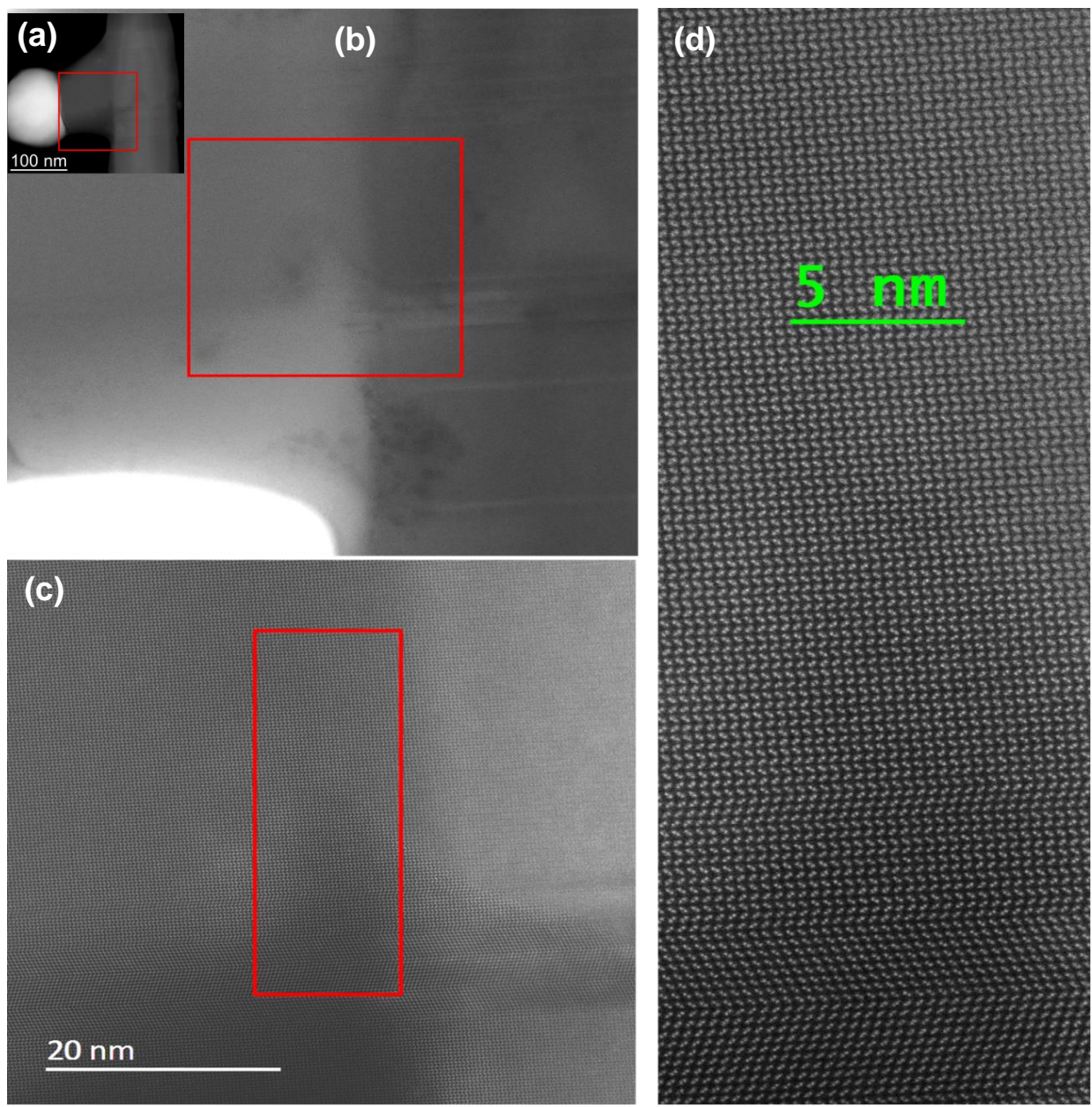

**Figure S6.** STEM image of the WZ GaAs NW branch generated at the WZ GaAs NW trunk (Sample 3). The branch finishes with the catalyzing side Bi droplet. Subsequent panels are magnifications of the regions marked with red rectangles.

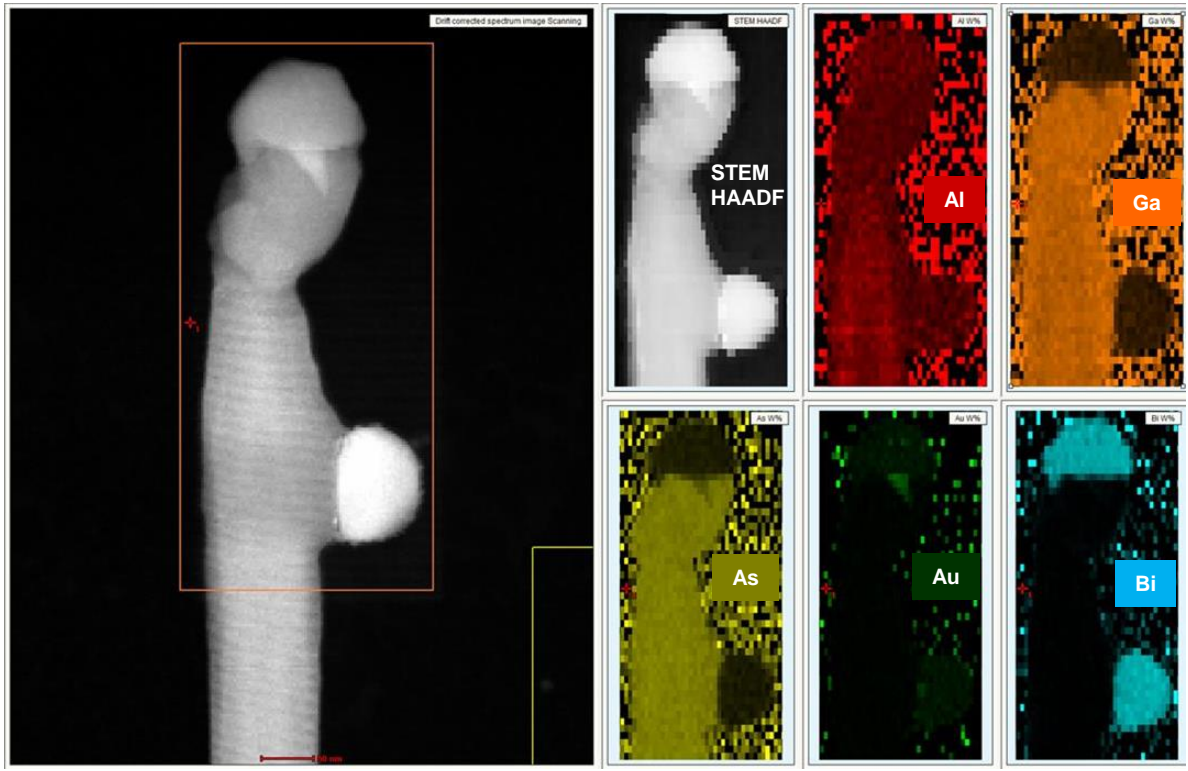

**Figure S7.** STEM image (left panel) and EDS composition maps (right panel) of the NW with preserved (unbroken) top part; collected from sample 3.

The STEM images and EDS composition maps of the upper part of the NW collected form sample 3, with unbroken top part are shown in Figure S7. The curved NW part below the very top Bi droplet is axially grown during the Ga(As,Bi) shell deposition and has purely ZB structure,

Figure S8 shows images corresponding to the thin cross-sections of a ZB part of NW selected form sample 3. The thickness of the analyzed NW cross-section is measured using position averaged convergent beam electron diffraction (PACBED) technique. This pattern is obtained using the 9.5 mrad converged electron beam scanned across the core area of the NW cross-section. The comparison of experimental and simulated PACBED patterns is very sensitive to

local thickness determination.[2] In our case the best match is measured as $l^2$-norm metric defined as

$$l^2(t) = \sqrt{\sum_{pixels}\left[\tilde{I}_{exp} - \tilde{I}_{sim}(t)\right]^2} \qquad (1)$$

Where $\tilde{I}_{exp}$ and $\tilde{I}_{sim}$ denote experimental and simulated PACBED patterns, and t denotes the sample thickness input to the simulation. The minimum difference is obtained for t ≅75 nm. With this specimen thickness only relatively larger composition fluctuations can be determined.

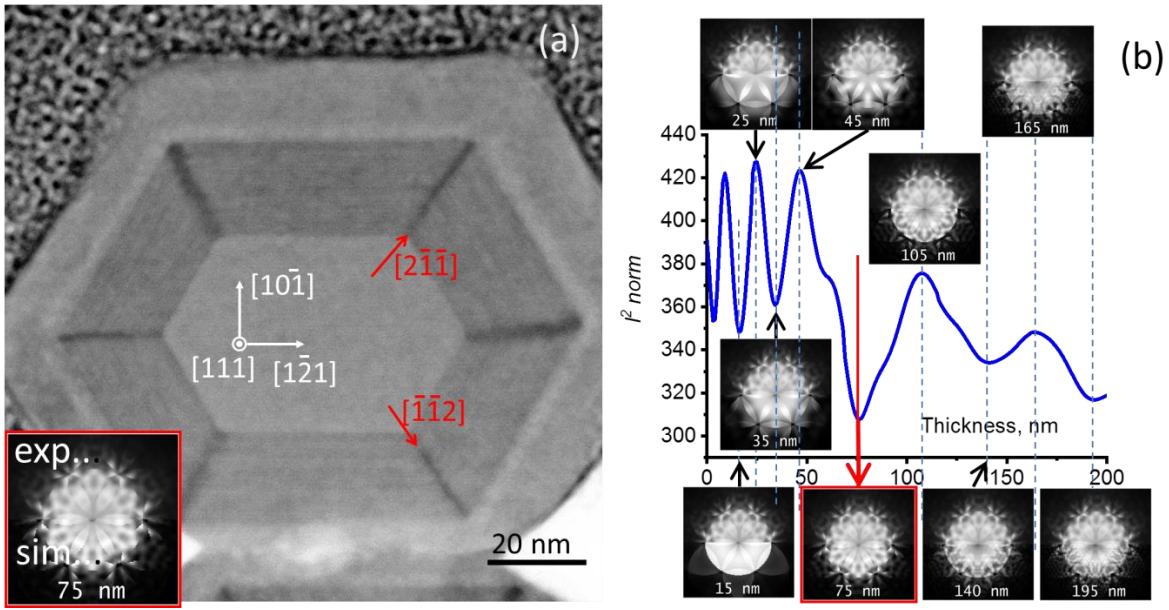

**Figure S8.** Analysis of a thin cross-section of ZB GaAs-(Ga,Al)As-Ga(As,Bi) NW segment (Sample 3). (a) HR-STEM image obtained with the camera length of 73 mm with corresponding best matched PACBED pattern inset; (b) the absolute difference between the experimental and simulated PACBED patterns measured as $l^2$-norm metric curve defined by Eq. (1) used for the thickness determination of FIB lamella; the minimum of $l^2$–norm is obtained for 75 nm thickness.